\documentclass[epj]{svjour}
\makeatletter
\def\cl@chapter{\@elt {theorem}}
\makeatother

\usepackage{graphics}
\usepackage{amsmath}
\usepackage{url}
\usepackage{cleveref}

\crefname{figure}{Fig.}{Fig.}

\newcommand{\tr}{\mathrm{Tr}}

\newcommand{\mat}[1]{#1}

\newcommand{\ypred}{\vec{y}_1}
\newcommand{\yobs}{\vec{y}_2}
\newcommand{\Kpred}{\mat{K}_{11}}
\newcommand{\Kobs}{\mat{K}_{22}}
\newcommand{\Kop}{\mat{K}_{21}}

\newcommand{\ypsi}{\vec{y}_3}
\newcommand{\Kpsi}{\mat{K}_{33}}
\newcommand{\Kobspsi}{\mat{K}_{23}}
\newcommand{\Kpsipred}{\mat{K}_{31}}
\newcommand{\Kobsapx}{\mat{\tilde{K}}_{22}}
\newcommand{\Kopapx}{\mat{\tilde{K}}_{21}}
\newcommand{\Kpredapx}{\mat{\tilde{K}_{11}}}

\newcommand{\Dpred}{\mat{D}_1}
\newcommand{\Dobs}{\mat{D}_2}

\begin{document}
\title{Estimating model bias over the complete nuclide chart with sparse Gaussian processes at the example of INCL/ABLA and double-differential neutron spectra}
\author{Georg Schnabel
}                     
\authorrunning{G. Schnabel}
\titlerunning{Estimating model bias with sparse Gaussian processes}
%
%
\institute{Irfu, CEA, Universit\'e Paris-Saclay, 91191 Gif-sur-Yvette, France}
\date{Received: date / Revised version: date}
%
\abstract{
Predictions of nuclear models guide the design of nuclear facilities to ensure their safe and efficient operation.
Because nuclear models often do not perfectly reproduce available experimental data, decisions based on their predictions may not be optimal.
Awareness about systematic deviations between models and experimental data helps to alleviate this problem.
This paper shows how a sparse approximation to Gaussian processes can be used to estimate the model bias over the complete nuclide chart at the example of inclusive double-differential neutron spectra for incident protons above 100\,MeV.
A powerful feature of the presented approach is the ability to predict the model bias for energies, angles, and isotopes where data are missing.
The number of experimental data points that can be taken into account is at least in the order of magnitude of~$10^4$ thanks to the sparse approximation.
The approach is applied to the Li\`ege Intranuclear Cascade Model (INCL) coupled to the evaporation code ABLA.
The results suggest that sparse Gaussian process regression is a viable candidate to perform global and quantitative assessments of models.
Limitations of a philosophical nature of this (and any other) approach are also discussed. 
} 
\maketitle
\section{Introduction}
\label{intro}
Despite theoretical advances, nuclear models are in general not able to reproduce all features of trustworthy experimental data.
Because experiments alone do not provide sufficient information to solve problems of nuclear engineering, models are still needed to fill the gaps.

Being in need of reliable nuclear data, a pragmatic solution to deal with imperfect  models is the introduction of a bias term on top of the model.
The true prediction is then given as the sum of model prediction and bias term.
The form of the bias term is in principle arbitrary and a low order polynomial could be a possible choice.
However, assumptions about how models deviate from reality are usually very vague, which makes it difficult to justify one parametrization over another one.

Methods of non-parametric statistics help to a certain extent to overcome this specification problem.
In particular, Gaussian process~(GP) regression (also known as Kriging, e.g.,~\cite{rasmussen_gaussian_2006}) enjoys popularity in various fields, such as geostatistics, remote sensing and robotics, due to its conceptual simplicity and sound embedding in Bayesian statistics.
Instead of providing a parameterized function to fit, one specifies a mean function and a covariance function.
The definition of these two quantities induces a prior probability distribution on a function space.
Several standard specifications of parametrized covariance functions exist, e.g.,~\cite[Ch.\ 4]{rasmussen_gaussian_2006}, whose parameters regulate the smoothness and the magnitude of variation of the function to be determined by GP regression.

The optimal choice of the values for these so-called hyperparameters is problem dependent and can also be automatically performed by methods such as marginal likelihood optimization and cross validation, e.g.,~\cite[Ch.\ 5]{rasmussen_gaussian_2006}.
In addition, features of various covariance functions can be combined by summing and multiplying them.
Both the automatic determination of hyperparameters and the combination of covariance functions will be demonstrated.

From an abstract viewpoint, GP regression is a method to learn a functional relationship based on samples of input-output associations without the need to specify a functional shape.
GP regression naturally yields besides estimates also the associated uncertainties.
This feature is essential for evaluating nuclear data, because uncertainties of estimates are as important for the design of nuclear facilities as are the estimates themselves.
Prior knowledge about the smoothness and the magnitude of the model bias can be taken into account.
Furthermore, many existing nuclear data
evaluation methods, e.g.,~\cite{muir_global_2007,leeb_consistent_2008,herman_development_2008},
can be regarded as special cases.
This suggests that the approach discussed in this paper can be combined with existing evaluation methods in a principled way.

The main hurdle for applying GP regression is the bad scalability in the number of data points~$N$.
The required inversion of an $N\times N$ covariance matrix leads to computational complexity of $N^3$.
This limits the application of standard GP regression to several thousand data points on contemporary desktop computers.
Scaling up GP regression to large datasets is therefore a field of active research.
Approaches usually rely on a combination of parallel computing and the introduction of sparsity in the covariance matrix, which means to replace the original covariance matrix by a low rank approximation, see e.g.,~\cite{quinonero-candela_unifying_2005}.

In this paper, I investigate the sparse approximation introduced in~\cite{snelson_sparse_2006} to estimate the model bias of inclusive double-differential neutron spectra over the complete nuclide chart for incident protons above 100\,MeV.
The predictions are computed by the C++ version of the Li\`ege Intranuclear Cascade Model (INCL)~\cite{mancusi_new_2014,mancusi_extension_2014} coupled to the Fortran version of the evaporation code ABLA07~\cite{abla07}.
The experimental data are taken from the EXFOR database~\cite{otuka_towards_2014}.

The idea of using Gaussian processes to capture deficiencies of a model exists for a long time in the literature, see e.g.,~\cite{blight_bayesian_1975}, and has also already been studied in the context of nuclear data evaluation, e.g.,~\cite{pigni_uncertainty_2003,schnabel_large_2015,schnabel_differential_2016}.
The novelty of this contribution is the application of GP regression to a large dataset with isotopes across the nuclide chart, which is possible thanks to the sparse approximation.
Furthermore, the way GP regression is applied enables predictions for isotopes without any data.

The exemplary application of sparse GP regression in this paper indicates that the inclusion of hundred thousands of data points may be feasible and isotope extrapolations    yield reasonable results if some conditions are met.
Therefore, sparse GP regression is a promising candidate to perform global assessments of models and to quantify their imperfections in a principled way.

%
%

The structure of this paper is as follows.
\Cref{sec:method} outlines the theory underlying sparse GP regression.
In particular, \Cref{sec:gpregression} provides a succinct exposition of standard GP regression, \cref{sec:gpregression} sketches how to construct a sparse approximation to a GP and how this approximation is exploited in the computation, and \cref{sec:marlikemax} explains the principle to adjust the hyperparameters of the covariance function based on the data.

The application to INCL/ABLA and inclusive double-differential neutron spectra is then discussed in~\cref{sec:application}.
After a brief introduction of INCL and ABLA in~\cref{sec:scenario}, the specific choice of covariance function is detailed in~\cref{sec:covfundesign}.
Some details about the hyperparameter adjustment are given in~\cref{sec:appmarlikemax} and the results of GP regression are shown and discussed in~\cref{sec:results}.

\section{Method}
\label{sec:method}

\subsection{GP regression}
\label{sec:gpregression}
GP regression, e.g.,~\cite{rasmussen_gaussian_2006}, can be derived in the framework of Bayesian statistics under the assumption that all probability distributions are multivariate normal.
Let the vector~$\ypred$ contain the values at the locations~$\{\vec{x}_i^p\}$ of interest and $\yobs$ the observed values at the locations~$\{\vec{x}_j^o\}$.

For instance, in the application in \cref{sec:application}, the elements in the vector~$\ypred$ represent the relative deviations of the ``truth" from the model predictions for neutron spectra at angles and energies of interest.
The vector~$\yobs$ contains the relative deviations of the available experimental data from the model predictions for neutron spectra at the angles and energies of the experiments.
The underlying assumption is that the experimental measurements differ from the truth by an amount compatible with their associated uncertainties.
If this assumption does not hold, model bias should be rather understood as a combination of model \textit{and} experimental bias.

Given a probabilistic relationship between the vectors $\ypred$ and $\yobs$, i.e. we know the conditional probability density function (pdf) of $\yobs$ given $\ypred$, $\rho(\yobs\,|\,\ypred)$ (e.g., because of continuity assumptions), the application of the Bayesian update formula yields
\begin{equation}
\rho(\ypred \,|\, \yobs) \propto \rho(\yobs \,|\, \ypred) \rho(\ypred) \,.
\end{equation}
The occurring pdfs are referred to as posterior $\rho(\ypred \,|\, \yobs)$, likelihood $\rho(\yobs \,|\, \ypred)$, and prior $\rho(\ypred)$.
The posterior pdf represents an improved state of knowledge.

The form of the pdfs in the Bayesian update formula can be derived from the joint distribution $\rho(\ypred,\yobs)$.
In the following, we need the multivariate normal distribution
\begin{multline}
\mathcal{N}(\vec{y}\,|\,\vec{\mu}, \mat{K}) =
\frac{1}{\sqrt{(2\pi)^N \det \mat{K}}} \times \\
\exp\left(
  -\frac{1}{2}
  (\vec{y} - \vec{\mu})^T 
  \mat{K}^{-1}
  (\vec{y} - \vec{\mu})
\right)
\end{multline}
which is characterized by the center vector~$\vec{\mu}$ and the covariance matrix~$\mat{K}$.
The dimension of the occurring vectors is denoted by~$N$.
Under the assumption that all pdfs are multivariate normal and centered at zero, the joint distribution of $\ypred$ and $\yobs$ can be written as
\begin{equation}
\rho(\ypred,\yobs) = \mathcal{N}\left(
\begin{pmatrix}
\ypred \\ \yobs
\end{pmatrix}
\,\middle| \,
  \begin{pmatrix} \vec{0} \\ \vec{0} \end{pmatrix},
  \begin{pmatrix}
  \Kpred & \Kop^T \\
  \Kop   & \Kobs 
  \end{pmatrix}
\right) \,.
\label{eq:blockmultvar}
\end{equation}
The compound covariance matrix contains the blocks $\Kpred$ and $\Kobs$ associated with $\ypred$ and $\yobs$, respectively.
The block $\Kop$ contains the covariances between the elements of $\ypred$ and $\yobs$.
Centering the multivariate normal pdf at zero is a reasonable choice for the estimation of model bias.
It means that an unbiased model is a priori regarded as the most likely option.
 
The posterior pdf is related to the joint distribution by 
\begin{equation}
\rho(\ypred \,|\, \yobs) = 
\frac{\rho(\ypred,\yobs)}
{\int \rho(\ypred,\yobs)\,d\ypred} \,.
\end{equation}
The solution for a multivariate normal pdf is another multivariate normal pdf.
For~\cref{eq:blockmultvar}, the result is given by
\begin{equation}
\rho(\ypred \,|\, \yobs) = \mathcal{N}(
\ypred \,|\, \ypred', \Kpred')
\end{equation}
with the posterior mean vector~$\ypred'$ and covariance matrix~$\Kpred'$ (e.g.,~\cite[A.\ 3]{rasmussen_gaussian_2006}),
\begin{align}
\ypred' &= \Kop^T \Kobs^{-1}\yobs \,, 
\label{eq:gppostmean}
\\
\Kpred' &= \Kpred - \Kop^T \Kobs^{-1} \Kop \,.
\label{eq:gppostcov}
\end{align}
The important property of these equations is the fact that the posterior moments depend only on the observed vector $\yobs$.
The introduction of new elements into $\ypred$ and associated columns and rows into $\Kpred$ in~\cref{eq:blockmultvar} has no impact on the already existing values in $\ypred'$ and $\Kpred'$.
In other words, one is not obliged to calculate all posterior expectations and covariances at once. They can be calculated sequentially.

GP regression is a method to learn a functional relationship $f(\vec{x})$ based on samples of input-output associations $\{(\vec{x}_1,f_1), (\vec{x}_2,f_2), \,\dots \}$.
The vector $\yobs$ introduced above then contains the observed functions values of $f(\vec{x})$, i.e. $\yobs = (f_1, f_2, \dots)^T$.
Assuming all prior expectations to be zero, the missing information to evaluate~\cref{eq:gppostmean,eq:gppostcov} are the covariance matrices.
Because predictions for $f(\vec{x})$ should be computable at all possible locations $\vec{x}$ and the same applies to observations, covariances between functions values must be available for all possible pairs $(\vec{x}_i, \vec{x}_j)$ of locations.
This requirement can be met by the introduction of a so-called covariance function $\kappa(\vec{x}_i,\vec{x}_j)$.
A popular choice is the squared exponential covariance function
\begin{equation}
\kappa(\vec{x}_i,\vec{x}_j) = \delta^2 \exp\left(
-\frac{(\vec{x}_i-\vec{x}_j)^2}{2\lambda^2} 
\right) \,.
\label{eq:examplecovfun}
\end{equation}
The parameter $\delta$ enables the incorporation of prior knowledge about the range functions values are expected to span.
The parameter $\lambda$ regulates the smoothness of the solution.
The larger $\lambda$ the slower the covariance function decays for increasing distance between $\vec{x}_1$ and $\vec{x}_2$ and consequently the more similar function values are at nearby locations.

If the set~$\{\vec{x}_i^p\}$ contains the locations of interest and~$\{\vec{x}_j^o\}$ the observed locations, the required covariance matrices in \cref{eq:gppostmean,eq:gppostcov} are given by
\begin{equation}
\label{eq:definitionKobs}
\Kobs = \begin{pmatrix}
\kappa(\vec{x}_1^o,\vec{x}_1^o) & \kappa(\vec{x}_1^o,\vec{x}_2^o) & \cdots \\
\kappa(\vec{x}_2^o,\vec{x}_1^o) & \kappa(\vec{x}_2^o,\vec{x}_2^o) & \cdots \\
\vdots & \vdots & \ddots
\end{pmatrix} 
\end{equation}
and 
\begin{equation}
\label{eq:definitionKop}
\Kop = \begin{pmatrix}
\kappa(\vec{x}_1^o,\vec{x}_1^p) & \kappa(\vec{x}_1^o,\vec{x}_2^p) & \cdots \\
\kappa(\vec{x}_2^o,\vec{x}_1^p) & \kappa(\vec{x}_2^o,\vec{x}_2^p) & \cdots \\
\vdots & \vdots & \ddots
\end{pmatrix} \,.
\end{equation}

\subsection{Sparse Gaussian processes}
\label{sec:sparseGP}
The main hurdle for applying GP regression using many observed pairs $(
\vec{x}_i, f_i)$ is the inversion of the covariance matrix $\Kobs$ in \cref{eq:gppostmean,eq:gppostcov}.
The time to invert this $N \times N$ matrix with $N$ being the number of observations increases proportional to $N^3$.
This limits the application of standard GP regression to several thousand observations on contemporary desktop computers.
Parallelization helps to a certain extent to push this limit.
Another measure is the approximation of the covariance matrix by a low rank approximation.
I adopted the sparse approximation described in~\cite{snelson_sparse_2006}, which will be briefly outlined here.
For a survey of different approaches and their connections consult e.g.,~\cite{quinonero-candela_unifying_2005}.

Suppose that we have not measured the values in $\yobs$ associated with the locations~$\{\vec{x}_j^o\}$, but instead a vector $\ypsi$ associated with some other locations $\{\vec{x}_k^{psi}\}$.
We refer to $\ypsi$ as vector of pseudo-inputs.
Now we use \cref{eq:gppostmean} to determine the hypothetical posterior expectation of $\yobs$,
\begin{equation}
\yobs' = \overbrace{\Kobspsi \Kpsi^{-1}}^{\mat{S}}\ypsi \,.
\label{eq:Kobsapx0}
\end{equation}
The matrices $\Kobspsi$ and $\Kpsi$ are constructed analogous to \cref{eq:definitionKop} and \cref{eq:definitionKobs}, respectively, i.e. $(\Kobspsi)_{ij} = \kappa(\vec{x}_i^o, \vec{x}_j^{psi})$ and $(\Kpsi)_{jk} = \kappa(\vec{x}_j^{psi}, \vec{x}_k^{psi})$ with $i=1..N, j=1..M$ and $N$ being the number of observations and $M$ the number of pseudo-inputs.
Noteworthy, $\yobs'$ is a linear function of $\ypsi$.
Under the assumption of a deterministic relationship, we can replace the posterior expectation $\yobs'$ by $\yobs$.
Using the sandwich formula and $\rho(\ypsi)=\mathcal{N}(\ypsi\,|\,\vec{0},\Kpsi)$, we get for the covariance matrix of $\yobs$
\begin{equation}
\Kobsapx = \mat{S}\Kpsi\mat{S}^T = \Kobspsi \Kpsi^{-1} \Kobspsi^T \,.
\label{eq:Kobsapx}
\end{equation}
Given that both $\Kpsi$ and $\Kobspsi$ have full rank and the number of observations $N$ is bigger than the number of pseudo-inputs $M$, the rank of $\Kobsapx$ equals $M$.
The approximation is more rigid than the original covariance matrix due to the lower rank.

In order to restore the flexibility of the original covariance matrix, the diagonal matrix $\Dobs=\mathop{diag}[\Kobs - \Kobsapx]$ is added to $\Kobsapx$.
This correction is essential for the determination of the pseudo-input locations via marginal likelihood optimization as explained in~\cite[Sec.\ 3]{snelson_sparse_2006}.
Furthermore, to make the approximation exhibit all properties of a GP, it is also necessary to add $\Dpred=\mathop{diag}[\Kpred - \Kpredapx]$ to $\Kpredapx = \Kpsipred^T \Kpsi^{-1} \Kpsipred$ as explained in~\cite[Sec.\ 6]{quinonero-candela_unifying_2005}.

Making the replacements $\Kobs \rightarrow \Kobsapx+\Dobs$, $\Kop \rightarrow \Kopapx = \Kobspsi \Kpsi^{-1} \Kpsipred$, and $\Kpred \rightarrow \Kpredapx + \Dpred$ in \cref{eq:gppostmean,eq:gppostcov}, we obtain
\begin{align}
\ypred' &= \Kopapx^T \left( \Kobsapx + \Dobs \right)^{-1} \yobs \,, 
\label{eq:gpapxpostmean}
\\
\Kpred' &= \Kpredapx + \Dpred - \Kopapx^T \left( \Kobsapx + \Dobs \right)^{-1} \Kopapx \,.
\label{eq:gpapxpostcov}
\end{align}
Using the Woodbury matrix identity (e.g.,~\cite[A.\ 3]{rasmussen_gaussian_2006}), these formulas can be rewritten as (e.g.,~\cite[Sec.\ 6]{quinonero-candela_unifying_2005})
\begin{align}
\ypred' &= \Kpsipred^T \left( \Kpsi + \Kobspsi^T \Dobs^{-1} \Kobspsi \right)^{-1}
\Kobspsi^T \mat{D}^{-1}  \yobs \,, 
\label{eq:gpapxpostmean1}
\\
\Kpred' &= \Dpred + \Kpsipred^T \left( \Kpsi + \Kobspsi^T \Dobs^{-1} \Kobspsi \right)^{-1} \Kpsipred \,.
\label{eq:gpapxpostcov1}
\end{align}
Noteworthy, the inversion in these expressions needs only to be performed for an $M \times M$~matrix where~$M$ is the number of pseudo-inputs.
Typically, $M$ is chosen to be in the order of magnitude of hundred.
The computational cost for inverting the diagonal matrix~$D$ is negligible.

Using \cref{eq:gpapxpostmean1,eq:gpapxpostcov1}, the computational complexity scales linearly with the number of observations~$N$ due to the multiplication by~$\Kobspsi$.
This feature enables to increase the number of observations by one or two orders of magnitude compared to \cref{eq:gppostmean,eq:gppostcov} in typical scenarios.

\subsection{Marginal likelihood maximization}
\label{sec:marlikemax}
Covariance functions depend on parameters (called hyperparamters) whose values must be specified.
In a full Bayesian treatment, the choice of values should not be informed by the same observations that enter into the GP regression afterwards.
In practice, this ideal is often difficult to achieve due to the scarcity of available data and therefore frequently abandoned.
A full Bayesian treatment may also become computationally intractable if there are too much data. 

Two popular approaches to determine the hyperparameters based on the data are marginal likelihood maximization and cross validation, see e.g.,~\cite[Ch.\ 5]{rasmussen_gaussian_2006}.
A general statement which approach performs better cannot be made.
I decided to use marginal likelihood optimization because it can probably be easier interfaced with existing nuclear data evaluation methods.

Given a covariance function depending on some hyperparameters, e.g., $\kappa(\vec{x}_i, \vec{x}_j2\,|\,\delta,\lambda)$ with hyperparameters $\delta$ and $\lambda$ as in \cref{eq:examplecovfun},
the idea of marginal likelihood maximization is to select values for the hyperparameters that maximize the probability density for the observation vector $\rho(\yobs)$.
In the case of the multivariate normal pdf in \cref{eq:blockmultvar}, it is given by (e.g.,~\cite[Sec.\ 5.4.1]{rasmussen_gaussian_2006})
\begin{equation}
\ln \rho(\yobs) = 
-\frac{N}{2}\log(2\pi) - \frac{1}{2}\log\det\Kobs 
- \frac{1}{2} \yobs^T \Kobs^{-1} \yobs
\label{eq:marlike}
\end{equation}
The first term is a constant, the second term is up to a constant the information entropy of the multivariate normal distribution, and the third term is the generalized $\chi^2$-value.
The maximization of this expression amounts to balancing two objectives: minimizing the information entropy and maximizing the $\chi^2$-value.

The partial derivative of \cref{eq:marlike} with respect to a hyperparameter is given by  (e.g.,~\cite[Sec.\ 5.4.1]{rasmussen_gaussian_2006})
\begin{multline}
\frac{\partial \ln \rho(\yobs)}{\partial \lambda} = 
-\frac{1}{2} \tr\left(
\Kobs^{-1} \frac{\partial\Kobs}{\partial \lambda}
\right) \\
+
\frac{1}{2} \yobs^T \Kobs^{-1} \frac{\partial\Kobs}{\partial \lambda} \Kobs^{-1} \yobs \,,
\label{eq:gradmarlike}
\end{multline}
which enables the usage of gradient-based optimization algorithms.
In this paper, I use the L-BFGS-B algorithm~\cite{byrd_limited_1995} because it can deal with a large number of parameters and allows to impose restrictions on their ranges.

Due to the appearance of the determinant and the inverse of $\Kobs$, the optimization is limited to several thousand observations on contemporary desktop computers.
However, replacing $\Kobs$ by the approximation $\Kobsapx + \Dobs$ in~\cref{eq:marlike,eq:gradmarlike} enables to scale up the number of observations by one or two orders of magnitude.
The structure of the approximation is exploited by making use of the matrix determinant lemma, the Woodbury identity, and the trace being invariant under cyclic permutations. 

The approximation to $\Kobs$ is not only determined by the hyperparameters but also by the location of the pseudo-inputs $\{\vec{x}_k^{psi}\}$.
Hyperparameters and pseudo-input locations can be jointly adjusted by marginal likelihood maximization.
The number of pseudo-inputs is usually significantly larger (e.g., hundreds) than the number of hyperparameters (e.g., dozens).
In addition, the pseudo-inputs are points in a potentially multi-dimensional space and their specification requires a coordinate value for each axis.
For instance, in \cref{sec:application} sparse GP regression is performed in a five dimensional space with three hundred pseudo-inputs, which gives 1500 associated parameters.
Because \cref{eq:gradmarlike} has to be evaluated for each parameter in each iteration of the optimization algorithm, its efficient computation is important.

The mathematical details are technical and tedious and hence only the key ingredient for efficient computation will be discussed.
Let $x_{kl}$ be the $l^\textrm{th}$ coordinate of the $k^\textrm{th}$ pseudo-input.
The crucial observation is that $\partial\Kpsi/\partial x_{kl}$ yields a matrix in which only the $l^\textrm{th}$ and $k^\textrm{th}$ column and row contain non-zero elements.
A similar statement holds for $\partial\Kobspsi/\partial x_{kl}$.
This feature can be exploited in the multiplications and the trace computation in~\cref{eq:gradmarlike} (where $\Kobs$ is substituted by $\Kobsapx+\Dobs$) to achieve $\mathcal{O}(N M)$ per coordinate of a pseudo-input with $M$ being the number of pseudo-inputs, and $N$ being the number of observations.
This is much more efficient than $\mathcal{O}(d N M^2)$ for the partial derivative with respect to a hyperparameter.

\section{Application}
\label{sec:application}

\subsection{Scenario}
\label{sec:scenario}
The model bias was determined for the C++ version of the Li\`ege Intranuclear Cascade Model (INCL)~\cite{mancusi_new_2014}, a Monte Carlo code, coupled to the evaporation code ABLA07~\cite{abla07} because this model combination performs very well according to an IAEA benchmark of spallation models~\cite{iaea_benchmark,david_spallation_2015} and is used in transport codes such as MCNPX and GEANT4.
This suggests that the dissemination of a more quantitative performance assessment of INCL coupled to ABLA potentially helps many people to make better informed decisions.
Because some model ingredients in INCL are based on views of classical physics (as opposed to quantum physics), the model is mainly used for high-energy reactions above 100\,MeV.

The ability of a model to accurately predict the production of neutrons and their kinematic properties may be regarded as one of the most essential features for nuclear engineering applications.
Especially for the development of the innovative research reactor MYRRHA~\cite{myrrha_reactor} driven by a proton accelerator, these quantities need to be well predicted for incident protons.

For these reasons, I applied the approach to determine the model bias in the prediction of inclusive double-differential neutron spectra for incident protons and included almost all nuclei for which I found data in the EXFOR database~\cite{otuka_towards_2014}.
Roughly ten thousand data points were taken into account.
\Cref{tbl:dataoverview} gives an overview of the data.


\begin{table}[ht]
\centering
\caption{Summary of the double-differential (p,X)n data above 100\,MeV incident energy found in EXFOR~\cite{otuka_towards_2014}.
No mass number is appended to the isotope name in the case of natural composition.
Columns are: incident proton energy (En), range of emitted neutron energy ($E_\textrm{min}$, $E_\textrm{max}$), range of emission angles ($\theta_\textrm{min}$, $\theta_\textrm{max}$), and number of data points (NumPts).
Energy related columns are in MeV and angles in degree.
The total number of data points is 9287.} 
\label{tbl:dataoverview}
\begin{tabular}{lrrrrrr}
  \hline
Isotope & En & $E_\textrm{min}$ & $E_\textrm{max}$ & $\theta_\textrm{min}$ & $\theta_\textrm{max}$ & NumPts \\ 
  \hline
C & 800 & 1.2 & 700 & 15 & 150 & 189 \\ 
  C & 1500 & 1.2 & 1250 & 15 & 150 & 245 \\ 
  C & 3000 & 1.2 & 2500 & 15 & 150 & 128 \\ 
  Na23 & 800 & 3.5 & 266 & 30 & 150 & 84 \\ 
  Al27 & 800 & 1.2 & 700 & 15 & 150 & 119 \\ 
  Al27 & 1000 & 2.5 & 280 & 15 & 150 & 223 \\ 
  Al27 & 1200 & 2.0 & 1189 & 10 & 160 & 404 \\ 
  Al27 & 1500 & 1.2 & 1250 & 15 & 150 & 129 \\ 
  Al27 & 1600 & 2.5 & 280 & 15 & 150 & 226 \\ 
  Al27 & 3000 & 1.2 & 2500 & 15 & 150 & 132 \\ 
  Fe & 800 & 1.2 & 771 & 10 & 160 & 505 \\ 
  Fe & 1200 & 2.0 & 1171 & 10 & 160 & 417 \\ 
  Fe & 1500 & 1.2 & 1250 & 15 & 150 & 129 \\ 
  Fe & 1600 & 2.0 & 1572 & 10 & 160 & 460 \\ 
  Fe & 3000 & 1.2 & 2500 & 15 & 150 & 133 \\ 
  Cu & 1000 & 2.5 & 280 & 15 & 150 & 227 \\ 
  Cu & 1600 & 2.5 & 280 & 15 & 150 & 231 \\ 
  Zr & 1000 & 2.5 & 280 & 15 & 150 & 229 \\ 
  Zr & 1200 & 2.0 & 1189 & 10 & 160 & 423 \\ 
  Zr & 1600 & 2.5 & 280 & 15 & 150 & 229 \\ 
  In & 800 & 1.2 & 700 & 15 & 150 & 116 \\ 
  In & 1500 & 1.2 & 1250 & 15 & 150 & 128 \\ 
  In & 3000 & 1.2 & 2500 & 15 & 150 & 133 \\ 
  W & 800 & 3.1 & 333 & 30 & 150 & 110 \\ 
  W & 1000 & 2.5 & 280 & 15 & 150 & 231 \\ 
  W & 1200 & 2.0 & 1189 & 10 & 160 & 413 \\ 
  W & 1600 & 2.5 & 280 & 15 & 150 & 231 \\ 
  Pb & 318 & 5.4 & 356 & 7 & 7 & 53 \\ 
  Pb & 800 & 1.2 & 771 & 10 & 160 & 624 \\ 
  Pb & 1000 & 2.5 & 280 & 15 & 150 & 231 \\ 
  Pb & 1200 & 2.0 & 1189 & 10 & 160 & 563 \\ 
  Pb & 1500 & 1.2 & 1250 & 15 & 150 & 249 \\ 
  Pb & 1600 & 2.0 & 1591 & 10 & 160 & 691 \\ 
  Pb & 3000 & 1.2 & 2500 & 15 & 150 & 131 \\ 
  Pb208 & 2000 & 0.4 & 402 & 30 & 150 & 170 \\ 
  Th232 & 1200 & 2.0 & 1189 & 10 & 160 & 351 \\ 
   \hline
\end{tabular}
\end{table}

\subsection{Design of the covariance function}
\label{sec:covfundesign}
The covariance function presented in~\cref{eq:examplecovfun} is probably too restrictive to be directly used on double-differential spectra.
It incorporates the assumption that the model bias spans about the same range for low and high emission energies.
Because the neutron spectrum quickly declines by orders of magnitude with increasing emission energy, it is reasonable to use a covariance function with more flexibility to disentangle the systematics of the model bias associated with these two energy domains.

Assuming the two covariance functions $\kappa_1(\vec{x}_i,\vec{x}_j)$ and $\kappa_2(\vec{x}_i,\vec{x}_j)$, a more flexible covariance function can be constructed in the following ways (e.g.,~\cite[Sec.\ 4.2.4]{rasmussen_gaussian_2006})
\begin{align}
\kappa_{1+2}(\vec{x}_i,\vec{x}_j) &= 
\kappa_1(\vec{x}_i,\vec{x}_j) +
\kappa_2(\vec{x}_i,\vec{x}_j) \,,
\label{eq:covfun_sum}
\\
\kappa_{1\times 2}(\vec{x}_i,\vec{x}_j) &= 
\kappa_1(\vec{x}_i,\vec{x}_j) \,
\kappa_2(\vec{x}_i,\vec{x}_j) \,.
\label{eq:covfun_prod}
\end{align}
Taking into account that the purpose of a covariance function is to compute elements of a covariance matrix, the construction in \cref{eq:covfun_sum} is analogous to the possible construction of an experimental covariance matrix:
An experimental covariance matrix can be assembled by adding a diagonal covariance matrix reflecting statistical uncertainties to another one with systematic error contributions.
Please note that sparse GP regression as presented in this paper works only with a diagonal experimental covariance matrix.
\Cref{eq:covfun_prod} will be used to achieve a transition between the low and high energy domains.

To state the full covariance function used in this paper, the following covariance function on a one-dimensional input space is introduced,
\begin{equation}
\kappa(x,x' \,|\, \lambda) = \exp\left(
-\frac{(x-x')^2}{2\lambda^2} 
\right) \,.
\label{eq:covfun1D}
\end{equation}
The meaning of the hyperparameter $\lambda$ was explained below \cref{eq:examplecovfun}.

The inclusive double-differential neutron spectra for incident protons over the complete nuclide chart can be thought of as a function of spectrum values associated with points in a five dimensional input space.
The coordinate axes are incident energy (En), mass number (A), charge number (Z), emission angle ($\theta$) and emission energy (E).
Using~\cref{eq:covfun1D} we define two covariance functions $\kappa_L$ and $\kappa_H$ associated with the low and high energy domain of the emitted neutrons.
Given two input vectors
\begin{align}
\vec{x}_i &= (\textrm{En}_i, A_i, Z_i, \theta_i, E_i)^T \,,  \\
\vec{x}_j &= (\textrm{En}_j, A_j, Z_j, \theta_j, E_j)^T \,,
\end{align}
the form of the covariance function for both $\kappa_L$ and $\kappa_H$ is assumed to be
\begin{multline}
\kappa_\mathcal{C}(\vec{x}_i,\vec{x}_j \,|\,\delta_\mathcal{C},\vec{\lambda}_\mathcal{C}) = 
\delta_\mathcal{C}^2 \,
\kappa(\textrm{En}_i,\textrm{En}_j \,|\, \lambda_\textrm{En}^\mathcal{C}) \\  \kappa(A_i,A_j \,|\, \lambda_A^\mathcal{C}) \, \kappa(Z_i,Z_j \,|\,  \lambda_Z^\mathcal{C}) \\
\kappa(\theta_i,\theta_j \,|\, \lambda_\theta^\mathcal{C}) \,
\kappa(E_i,E_j \,|\, \lambda_E^\mathcal{C}) \,.
\end{multline}
The hyperparameters $\lambda_x^\mathcal{C}$ are for the coordinate axis indicated by $x \in \{\textrm{En},A,\dots\}$ and can take different values for $\kappa_L$ and $\kappa_H$ ($\mathcal{C}\in\{L,H\})$.

The transition between the two energy domains, which means to switch from $\kappa_L$ to $\kappa_H$, is established by the logistic function
\begin{equation}
\sigma(\vec{x} \,|\,\vec{k},\vec{x}_0) = 
\frac{1}{1+\exp(-\vec{k} (\vec{x}-\vec{x}_0))} \,.
\label{eq:logisticfun}
\end{equation}
Noteworthy, all the variables are vectors.
The equation $\vec{k} (\vec{x}-\vec{x}_0)=\vec{0}$ defines a (hyper)plane with normal vector $\vec{k}$ and distance $|\vec{k}\vec{x}_0|/|\vec{k}|$ to the origin of the coordinate system.
The function in \cref{eq:logisticfun} attains values close to zero for $\vec{x}$ far away from the plane on one side and close to one if far away on the other side.
Within which distance to the plane the transition from zero to one occurs depends on the length of $\vec{k}$. 
The larger $|\vec{k}|$, the faster the transition and the narrower the window of transition around the plane.

With the abbreviations
\begin{align}
\tau_L(\vec{x}_i,\vec{x}_j\,|\,\vec{k},\vec{x}_0) = \, &  
\sigma(\vec{x}_i \,|\,\vec{k},\vec{x}_0)
\sigma(\vec{x}_j \,|\,\vec{k},\vec{x}_0) \,, \\
\tau_H(\vec{x}_i,\vec{x}_j\,|\,\vec{k},\vec{x}_0) = \, & 
(1 - \sigma(\vec{x}_i \,|\,\vec{k},\vec{x}_0)) \times \\
& (1-\sigma(\vec{x}_j \,|\,\vec{k},\vec{x}_0)) \,, 
\end{align}
the full covariance function $\kappa_\textrm{full}$ is given by
\begin{multline}
\kappa_\textrm{full}(\vec{x}_i,\vec{x}_j) =
\tau_L(\vec{x}_i,\vec{x}_j\,|\,\vec{k},\vec{x}_0) \,
\kappa_L(\vec{x}_i,\vec{x}_j \,|\,\delta_L,\vec{\lambda}_H) \\
+ \tau_H(\vec{x}_i,\vec{x}_j\,|\,\vec{k},\vec{x}_0) \,
\kappa_H(\vec{x}_i,\vec{x}_j \,|\,\delta_H,\vec{\lambda}_H) \,.
\label{eq:fullcovfun}
\end{multline}

Finally, the GP regression is not performed on the absolute difference $\Delta$ between a model prediction $\sigma_\textrm{mod}$ and experimental data point $\sigma_\textrm{exp}$, but on the transformed quantity
\begin{equation}
\label{eq:trafomodelbias}
\tilde{\Delta} = 
\frac{(\sigma_\textrm{exp} - \sigma_\textrm{mod})}
{\max(\sigma_\textrm{mod}, 0.1)} \,.
\end{equation}
In words, relative differences are taken for model predictions larger than 0.1 and absolute differences scaled up by a factor of ten for model predictions below 0.1.
Relative differences fluctuate usually wildly for spectrum values close to zero---especially for a Monte Carlo code such as INCL---and the switch to absolute values helps  GP regression to find more meaningful solutions with a better ability to extrapolate.

Due to the number of roughly ten thousand data points, the covariance matrices computed with~\cref{eq:fullcovfun} were replaced by the sparse approximation outlined in~\cref{sec:sparseGP}.
I introduced three hundred pseudo-inputs and placed them randomly at the locations of the experimental data.
Their locations were then jointly optimized with the hyperparameters, which will be discussed in~\cref{sec:appmarlikemax}.

The diagonal matrix $\mat{D}_2$ occurring in the approximation was changed to
\begin{equation}
\tilde{\mat{D}}_2 = \mat{D}_2 + \mat{B} + \mat{P} \,,
\end{equation}
to accommodate statistical uncertainties of the model prediction (due to INCL being a Monte Carlo code) and the experimental data. 
Both $\mat{B}$ and $\mat{P}$ are diagonal matrices.
The matrix $\mat{B}$ contains variances corresponding to $10\%$ statistical uncertainty for all experimental data points.
The matrix $\mat{P}$ contains the estimated variances of the model predictions.

A diagonal matrix for the experimental covariance matrix $\mat{B}$ can certainly be challenged because the important systematic errors of experiments reflected in off-diagonal elements are neglected.
This is at the moment a limitation of the approach.

\subsection{Marginal likelihood maximization}
\label{sec:appmarlikemax}
The hyperparameters appearing in~\cref{eq:fullcovfun} and the locations of the three hundred pseudo-inputs~$\{\vec{x}_k^{psi}\}$ determining the approximation in~\cref{eq:Kobsapx0} were adjusted via marginal likelihood maximization described in~\cref{sec:marlikemax}.
To be explicit, the hyperparameters considered were $\delta_L$, $\delta_H$,
\begin{align}
\vec{\lambda}_L &= (\lambda_\textrm{En}^L, \lambda_A^L, \lambda_Z^L, \lambda_\theta^L,  \delta_E^L) \,, \\
 \vec{\lambda}_H &= (\lambda_\textrm{En}^H, \lambda_A^H, \lambda_Z^H, \lambda_\theta^H,
 \delta_E^H) \,,
\end{align}
and also $\vec{x}_0$ and $\vec{k}$ of the logistic function.
The vector $\vec{k}$ was forced to remain parallel to the axes associated with $\textrm{En}$, $A$, and $Z$, i.e. $\vec{k}=(0,0,0,k_\theta,k_E)^T$.
Further, polar coordinates $k_\theta=k_c\sin\gamma, k_E=k_c\cos \gamma$ were introduced.
The direction of the vector $\vec{x}_0$ was taken equal to that of $\vec{k}$, which removes ambiguity in the plane specification without shrinking the set of possible solutions.
Because of these measures, it sufficed to consider the length $x_c=|\vec{x}_0|$ as hyperparameter. 
Counting both hyperparameters and pseudo-inputs, 1515 parameters were taken into account in the optimization.

\begin{figure}[t]
  \centering
  \caption{Dispersion of the pseudo-inputs during the optimization procedure shown in the (A,Z)-projection.
  The black points are the initial positions and the green points the final positions of the pseudo-inputs.}
  \label{fig:pseudoinpevo}
  \includegraphics{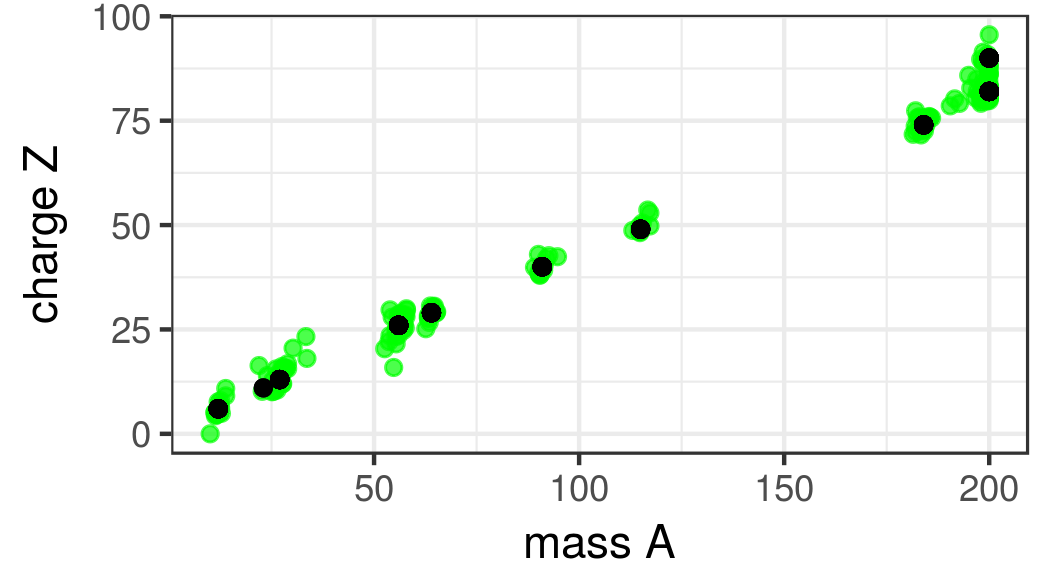}
\end{figure}

I employed the L-BFGS-B algorithm~\cite{byrd_limited_1995} as implemented in the \textit{optim} function of \textit{R}~\cite{r_development_core_team_r_2008}, which makes use of an analytic gradient, can deal with a large number of variables and permits the specification of range restrictions.
The optimization was performed on a cluster using 25 cores and was stopped after 3500 iterations, which took about 10 hours.
The obtained solution corresponds to $\chi^2/N = 1.03$ and is with a two-sided p-value of $0.04$  reasonably consistent in a statistical sense.
Restrictions of parameter ranges were established to introduce prior knowledge and to guide the optimization procedure.
Noteworthy, lower limits on length-scales, such as $\lambda_\theta^L$ and $\lambda_E^L$; were introduced to counteract their dramatic reduction due to inconsistent experimental data in the same energy/angle range.
\Cref{tbl:optimsetup} summarizes the optimization procedure.
The evolution of the pseudo-inputs projected onto the (A,Z)-plane is visualized in~\cref{fig:pseudoinpevo}.

A thorough study of the optimization process exceeds the scope of this paper and hence I content myself with a few remarks.
The length scales associated with the emission angle, i.e. $\lambda_\theta^\mathcal{C}$, experienced significant changes.
Their increase means that the model bias is similar for emission angles far away from each other and the GP process is able to capture them.
Concerning the length scales associated with the emission energy, the small value $\lambda_E^L=4.8$ compared to the larger value $\lambda_E^H=43$ indicates that the features of the model bias of the low and high energy domain are indeed different.
The most striking feature, however, is that the large length scales $\lambda_A^C$ and $\lambda_Z^C$ "survived" the optimization, which means that the model bias behaves similar over large regions of the nuclide charts.
Examples of isotope extrapolations will be given in~\cref{sec:results}.

As a final remark, a more rigorous study of the optimization procedure is certainly necessary and there is room for improvement.
This is left as future work.

\begin{table}[t]
\centering
\caption{Setup of the L-BFGS-B optimization and evolution of hyperparameters.
The columns LB and UB give the lower bound and upper bound, respectively, of the parameter ranges.
Columns It$_i$ indicate the parameter values after $i\times 1000$ iterations.
The column It$_0$ contains initial values and It$_f$ the final values after 3500 iterations.} 
\label{tbl:optimsetup}
\begin{tabular}{crrrrrrr}
  \hline
Name & LB & UB & It$_0$ & It$_1$ & It$_2$ & It$_3$ & It$_f$ \\ 
  \hline
$\delta_L$ & 0.01 & 0.50 & 0.05 & 0.50 & 0.50 & 0.50 & 0.50 \\ 
  $\lambda_\textrm{En}^L$ & 50 & 1000 & 100 & 100 & 99 & 99 & 99 \\ 
  $\lambda_A^L$ & 10 & 300 & 100 & 100 & 101 & 103 & 103 \\ 
  $\lambda_Z^L$ & 10 & 200 & 40 & 39 & 39 & 40 & 41 \\ 
  $\lambda_\theta^L$ & 10 & 179 & 50 & 51 & 57 & 66 & 68 \\ 
  $\lambda_E^L$ & 2 & 1000 & 20.0 & 7.3 & 5.0 & 4.7 & 4.8 \\ 
  $\delta_H$ & 0.01 & 0.50 & 0.40 & 0.33 & 0.34 & 0.33 & 0.33 \\ 
  $\lambda_\textrm{En}^H$ & 50 & 1000 & 300 & 292 & 285 & 275 & 272 \\ 
  $\lambda_\textrm{A}^H$ & 10 & 300 & 100 & 107 & 110 & 114 & 115 \\ 
  $\lambda_\textrm{Z}^H$ & 10 & 200 & 40 & 49 & 50 & 50 & 49 \\ 
  $\lambda_\theta^H$ & 10 & 179 & 50 & 62 & 63 & 64 & 64 \\ 
  $\lambda_E^H$ & 10 & 1000 & 20 & 40 & 41 & 42 & 43 \\ 
  $k_c$ & 0.1 & 20 & 0.2 & 0.4 & 0.4 & 0.3 & 0.3 \\ 
  $x_c$ & 2.0 & 500 & 6.0 & 2.6 & 2.3 & 2.4 & 2.8 \\ 
  $\gamma$ & 1.0 & 2.1 & 1.6 & 1.6 & 1.6 & 1.6 & 1.6 \\ 
   \hline
\end{tabular}
\end{table}

\subsection{Results and discussion}
\label{sec:results}
The values of the hyperparameters and pseudo-inputs obtained by marginal likelihood maximization were used in the covariance function in~\cref{eq:fullcovfun}.
This covariance function was employed to compute the required covariance matrices in~\cref{eq:gpapxpostmean1,eq:gpapxpostcov1} based on the experimental data summarized in~\cref{tbl:dataoverview}.
Because the hyperparameters are determined during hyperparameter optimization before being used in the GP regression, they will be referred to as prior knowledge.

\begin{figure*}[t]
\caption{Model bias of INCL in terms of \cref{eq:trafomodelbias} in the (p,X)n double differential spectra for 800 MeV incident protons and different isotopes as predicted by GP regression.
A missing mass number behind the isotope symbol indicates natural composition.
The uncertainty band of the prediction and the error bars of the experimental data denote the 2$\sigma$ confidence interval.
Carbon and indium were taken into account in the GP regression but not cadmium and oxygen.
The experimental data are colored according to the associated access number (ACCNUM) in the EXFOR database.
This shows that all displayed data come from just three experiments~\cite{amian_differential_1992,nakamoto_spallation_1995,ishibashi_measurement_1997}.
}
\label{fig:modelbiasexamples}       
\includegraphics{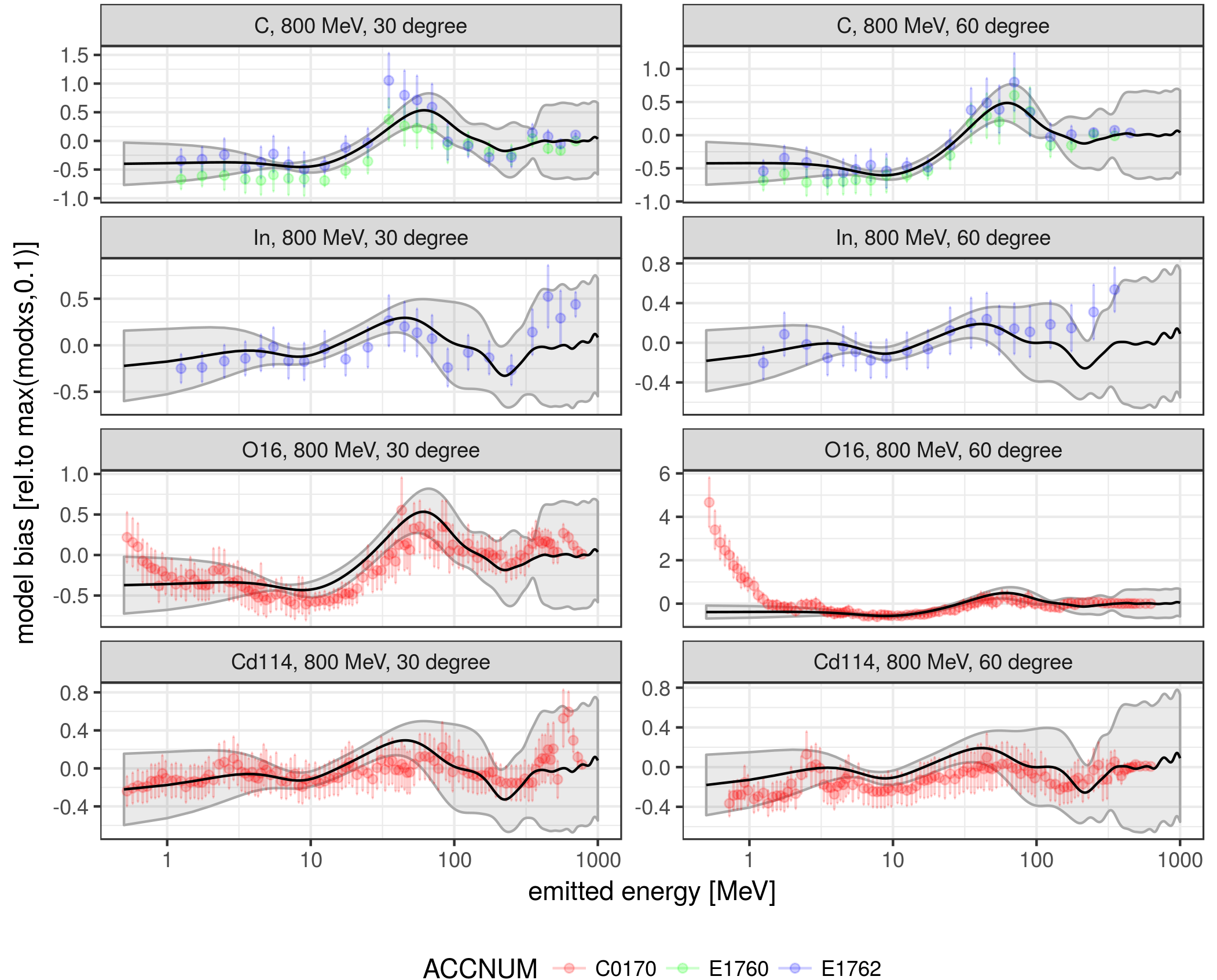}
\end{figure*}

\Cref{eq:gpapxpostmean1,eq:gpapxpostcov1} enable the prediction of a plethora of spectrum values and their uncertainties for combinations of incident energy, mass number, charge number, emission angle, and emission energy.
The few selected examples of predictions in~\cref{fig:modelbiasexamples} serve as the basis to discuss general features of GP regression, the underlying assumptions, and its accuracy and validity.

How well we can interpolate between data and how far we can extrapolate beyond the data depends on the suitability of the covariance function for the problem at hand.
The building block for the full covariance function in~\cref{eq:fullcovfun} is the one-dimensional covariance function in~\cref{eq:covfun1D}.
Using the latter imposes the assumption that possible solutions have derivatives of any order and hence are very smooth~\cite[Ch.\ 4]{rasmussen_gaussian_2006}.
Interpolations between data points included in the regression are determined by this smoothness property and values of the length scales~$\lambda_x^\mathcal{C}$.

The length scales reflect the prior assumption about similarity between spectrum values of points a certain distance away from each other.
This prior assumption directly impacts the uncertainty of the predictions.
The farther away a prediction is from an observation point, the higher the associated uncertainty.
If a prediction is already multiples of any length scale away from all observations, the uncertainty reverts to its prior value given by either $\delta_L$ or $\delta_H$ depending on the energy domain.

In the case of the sparse approximation, the uncertainty is related to the distance to the pseudo-inputs.
Because only few pseudo-inputs are located at very high and very low emission energies, the $2\sigma$ uncertainty bands in~\cref{fig:modelbiasexamples} in those energy domains are rather large despite the presence of experimental data.

The important finding in this specific application is that the length scales related to emission angle, $\lambda_E^\mathcal{C}$, mass number,  $\lambda_A^\mathcal{C}$, and charge number, $\lambda_E^\mathcal{C}$, are very large.
GP regression is therefore able to interpolate and extrapolate over large ranges of these axes.

The interpolation of spectrum values between angles and emission energies of an isotope with available data may be considered rather standard.
For instance, one can do a $\chi^2$-fit of a low order Legendre polynomial to the angular distributions for emission energies with data.
The coefficients of the Legendre polynomial for intermediate emission energies without data can then be obtained by linear interpolation.

The important difference between such a conventional fit and GP regression is the number of basis functions.
Whereas their number is fixed and limited in a conventional fit, GP regression amounts to a fit with an infinite number of basis functions~\cite[Sec.\ 2.2]{rasmussen_gaussian_2006}.
The length scales regulate the number of basis functions that effectively contribute to the solution.
Hyperparameter optimization decreases the length scales for unpredictable data which leads to a greater number of contributing basis functions and consequently to greater flexibility and larger uncertainties in the predictions.
This feature sets GP regression apart from a standard $\chi^2$-fit. In the latter, the uncertainty of the solution depends to a much lesser extent on the (un)predictability of the data and much more on the number of data points and the fixed number of basis functions.

One truly novel element in the approach is the inclusion of the mass and charge number, which enables predictions for isotopes without data.
We can easily imagine that different isotopes differ significantly by their physical properties.
From this perspective, the idea to extrapolate the model bias to other isotopes should be met with healthy skepticism.

To get a first grasp on the validity of isotope extrapolations, let us consider again the hyperparameter optimization discussed in~\ref{sec:appmarlikemax}.
The hyperparameters were adjusted on the basis of the isotopes in~\cref{tbl:dataoverview}.
These data are spread out over the periodic table and cover a range from carbon to thorium.
In these data, similar trends of the model bias persist across significant ranges of the mass and charge number, which was the reason that the associated length scales retained high values during optimization.
For instance, the experimental data of carbon and indium in \cref{fig:modelbiasexamples}  show comparable structures of the model bias despite their mass differences.
 
However, the isotopes considered in the optimization are not very exotic and gaps between them are at times large.
Further, these isotopes are certainly not a random sample taken from the periodic table and therefore most theoretical guarantees coming from estimation theory do not hold.
So how confident can we be about isotope extrapolation?

To provide a basis for the discussion of this question, \cref{fig:modelbiasexamples} also contains predictions for oxygen and cadmium.
Importantly, the associated experimental data have not been used for the hyperparameter optimization and in the GP regression. 
The predictions follow well the trend of the experimental data.
The supposedly $2\sigma$ uncertainty bands, however, include less than $95\%$ of the data points. 
This observation points out that there are systematic differences between isotopes.
Therefore, due to the sample of isotopes not being a random sample, uncertainty bands should be interpreted with caution.

One way to alleviate this issue could be to add a covariance function to~\cref{eq:fullcovfun} which only correlates spectrum values of the same isotope but not between isotopes.
This measure would lead to an additional uncertainty component for each isotope, which only decreases if associated data are included in the GP regression.

A very extreme mismatch up to $500\%$ between the prediction and experimental data occurs for oxygen at 60$^\circ$ and emission energies below 1\,MeV.
The creation of low energy neutrons is governed by de-excitation processes of the nucleus suggesting that the angular distribution of emitted particles is isotropic.
This property holds for the model predictions but not for the experimental data.
The experimental data exhibits a peak at 60$^\circ$ which is about a factor five higher than at 30$^\circ$, 120$^\circ$ and 150$^\circ$.
The origin of this peculiarity may deserve investigation but is outside the scope of this work. For the sake of argument I assume that it is indeed a reflection of some property of the nucleus.

Because the data in~\cref{tbl:dataoverview} do not include any measurements below 1\,MeV emission energy in this mass range, both hyperparameter optimization and GP regression had no chance to be informed about such a feature.
In this case, the predictions and uncertainties are determined by nearby values or---if there are not any---by the prior expectation.

If we would have been aware of such effects, we could have used it as a component in the covariance function to reflect our large uncertainty about the spectrum for low emission energies.
Otherwise, the only sensible data-driven way to provide predictions and uncertainties for unobserved domains is to assume that effects are similar to those in some observed domains.
Of course, it is our decision which domains are considered similar.
The employed covariance function in~\cref{eq:fullcovfun} incorporates the assumption that nearby domains in terms of mass charge, angle, etc. are similar.
However, we are free to use any other assumption about similarity to construct the covariance function.
As a side note, also results of other models relying on different assumptions could serve as a basis for uncertainties in unobserved regions.
Such information can be included in GP regression in a principled way.

\section{Summary and Outlook}
\label{sec:summary}
Sparse GP regression, a non-parametric estimation technique of statistics, was employed to estimate the model bias of the C++ version of the Li\`ege Intranuclear Cascade Model (INCL) coupled to the evaporation code ABLA07.
Specifically, the model bias in the prediction of double-differential inclusive neutron spectra over the complete nuclide chart was investigated for incident protons above 100\,MeV. 
Experimental data from the EXFOR database served as the basis of this assessment.
Roughly ten thousand data points were taken into account.
The hyperparameter optimization was done on a computing cluster whereas the GP regression itself on a desktop computer.
The obtained timings indicate that increasing the number of data points by a factor of ten could be feasible.

For this specific application, it was shown that GP regression produces reasonable results for isotopes that have been included in the procedure.
It was argued that the validity of predictions and uncertainties for isotopes not used in the procedure depends on the validity of the assumptions made about similarity between isotopes.
As a simple benchmark, the isotopes oxygen and cadmium, which have not been taken into account in the procedure, were compared to the respective predictions.
The agreement between prediction and experimental data was reasonable but the 95\% confidence band sometimes misleading and should therefore be interpreted with caution.
Accepting the low energy peak of oxgyen at 60$^\circ$ in the data as physical reality,
the low energy spectrum of oxygen served as an example where the similarity assumption between isotopes did not hold.

As for any other uncertainty quantification method, it is a hard if not impossible task to properly take into account unobserved phenomena without any systematic relationship to observed ones.
The existence of such phenomena are \textit{unknown unknowns}, their observation is tagged shock, surprise or discovery, and they potentially have a huge impact where they appear. 

Even though GP regression cannot solve the philosophical problem associated with unknown unknowns, knowledge coming from the observation of new effects can be taken into account by modifying the covariance function.
For instance, the peculiar peak in the oxygen spectrum suggests the introduction of a term in the covariance function which increases the uncertainty of the spectrum values in the low emission energy domain. 
The ability to counter new observations with clearly interpretable modifications of the covariance function represents a principled and transparent way of knowledge acquisition.

The scalability and the possibility to incorporate prior assumptions by modeling the covariance function makes sparse GP regression a promising candidate for global assessments of nuclear models and to quantify their uncertainties.
Because the formulas of GP regression are structurally identical to those of many existing evaluation methods, GP regression should be regarded more as a complement than a substitute, which can be interfaced with existing evaluation methods.

\section{Acknowledgments}
This work was performed within the work package WP11 of the CHANDA project (605203) financed by the European Commission.
Thanks to Sylvie Leray, Jean-Christophe David and Davide Mancusi for useful discussion.

\bibliographystyle{epj}
\bibliography{bibliography}

\end{document}